\documentclass[aps, preprint, groupedaddress, floatfix, longbibliography]{revtex4-2}

\usepackage{epsfig}
\usepackage{epstopdf}
\usepackage{graphicx}
\usepackage{amsmath}
\usepackage{color}
\usepackage[export]{adjustbox}
\usepackage[colorlinks=true, allcolors=blue]{hyperref}
\usepackage{lineno}
\begin{document}

\title{Sensitive dependence of pairing symmetry on Ni-$e_g$ crystal field splitting in the nickelate superconductor La$_3$Ni$_2$O$_7$}

\author{Chengliang Xia$^1$\footnote{These authors contributed equally to this work.}, Hongquan Liu$^{1,2}$\footnotemark[1], Shengjie Zhou$^1$ and Hanghui Chen$^{1,3}$\footnote{hanghui.chen@nyu.edu}}
\affiliation{$^1$NYU-ECNU Institute of Physics, NYU Shanghai, Shanghai 200122, China\\
  $^2$Department of Physics, Brown University, 182 Hope Street, Providence, RI 02912, USA\\
  $^3$Department of Physics, New York University, New York, New York 10012, USA}

\date{\today}

\begin{abstract}
  The discovery of high-temperature superconductivity in La$_3$Ni$_2$O$_7$ under pressure has drawn great attention. However, consensus has not been reached on its pairing symmetry in theory. By combining density-functional-theory (DFT), maximally-localized-Wannier-function, and linearized gap equation with random-phase-approximation, we find that the pairing symmetry of La$_3$Ni$_2$O$_7$ is $d_{xy}$, if its DFT band structure is accurately reproduced by a downfolded bilayer two-orbital model. More importantly, we reveal that the pairing symmetry of La$_3$Ni$_2$O$_7$ sensitively depends on the crystal field splitting between two Ni-$e_g$ orbitals. A slight increase in Ni-$e_g$ crystal field splitting alters the pairing symmetry from $d_{xy}$ to $s_{\pm}$. Such a transition is associated with the change in inverse Fermi velocity and susceptibility, while the shape of Fermi surface remains almost unchanged. Our work highlights the sensitive dependence of pairing symmetry on low-energy electronic structure in multi-orbital superconductors, which calls for care in the downfolding procedure when one calculates their pairing symmetry.
\end{abstract}

\maketitle
\newpage

\section{Introduction}

The recent discovery of superconductivity in La$_3$Ni$_2$O$_7$ under pressure has drawn great attention~\cite{327nature,Hou_2023,PhysRevB.108.125105,PhysRevLett.131.126001,PhysRevB.108.L140504,PhysRevB.108.174501,PhysRevB.108.165141,zhang2023structural,liu2023spmwave,PhysRevB.108.L140505,qu2023bilayer,gu2023effective,fan2023superconductivity,yang2023interlayer,luo2023hightc,PhysRevB.109.104508,PhysRevB.108.L201121,PhysRevB.109.L180502,jiang2023high,wang2024electronic,PhysRevB.108.214522,PhysRevB.108.174511,geisler2023structural,PhysRevB.109.045127,labollita2023electronic,zhang2023hightemperature,yang2023orbitaldependent,pan2023effect,PhysRevB.109.165154,zhang2023electronic,ryee2023critical,lu2023interplay,wang2023pressureinduced,yang2023strong,kaneko2023pair,PhysRevMaterials.8.044801,zhang2023strong,PhysRevLett.132.146002,chen2023critical,cao2023flat,christiansson2023correlated,shen2023effective,li2023signature,chen2023orbitalselective,wang2023structure,wang2023observation,zhu2024superconductivity,schlomer2023superconductivity,zhang2023effects,lu2023superconductivity,wú2023charge,liu2023electronic,zhang2023electronic2,li2024pressuredriven}. It not only adds a new member to the family of nickelate superconductors~\cite{li2019superconductivity,PhysRevLett.125.027001,PhysRevLett.125.147003,gu2020single,gu2020substantial,doi:10.1126/sciadv.abl9927,https://doi.org/10.1002/adma.202104083,PhysRevB.105.115134,chen2023electronic}, but also substantially increases the transition temperature to near 80 K~\cite{327nature}. Under a pressure above 19 GPa, La$_3$Ni$_2$O$_7$ transforms from an orthorhombic structure (space group $Amam$) to a tetragonal structure (space group $I4/mmm$) at low temperatures~\cite{wang2023structure,li2024pressuredriven}. Extensive density-functional-theory (DFT) calculations find that the Fermi surface of tetragonal La$_3$Ni$_2$O$_7$ has three Fermi sheets that are derived from two Ni-$e_g$ orbitals~\cite{PhysRevLett.131.126001,liu2023spmwave,PhysRevB.108.165141,zhang2023structural}. This is different from the well-known single-orbital Fermi surface of cuprate superconductors~\cite{RevModPhys.72.969,keimer2015quantum,PhysRevLett.110.216405}. Based on this low-energy electronic structure, a number of studies use many-body methods and propose different mechanisms for the superconductivity found in La$_3$Ni$_2$O$_7$. Yet no consensus has been reached so far on the most favorable pairing symmetry of La$_3$Ni$_2$O$_7$--a central property of unconventional superconductivity. Refs.~\cite{gu2023effective,zhang2023structural,PhysRevB.108.165141,PhysRevB.108.L140505,PhysRevB.108.174501,qu2023bilayer,liu2023spmwave,PhysRevB.108.L140504,PhysRevB.109.165154} find $s_{\pm}$ symmetry (group theory notation $A_{1g}$), reminiscent of iron-based superconductors~\cite{PhysRevB.78.134512,Hirschfeld_2011,RevModPhys.83.1589}, while Refs.~\cite{PhysRevB.108.L201121,PhysRevB.109.104508,jiang2023high,PhysRevB.109.L180502,wang2024electronic,fan2023superconductivity} obtain $d_{xy}$ ($B_{2g}$) or $d_{x^2-y^2}$ ($B_{1g}$) symmetry.



In this work, we combine density-functional-theory (DFT)~\cite{PhysRev.136.B864,PhysRev.140.A1133}, maximally-localized-Wannier-function (MLWF)~\cite{RevModPhys.84.1419,MOSTOFI2008685}, and linearized gap equation calculations with random-phase-approximation (RPA)~\cite{PhysRevB.31.4403,YANASE20031,Graser_2009,RevModPhys.84.1383,PhysRevB.101.060504,PhysRevLett.123.247001} to study the superconducting pairing symmetry of tetragonal La$_3$Ni$_2$O$_7$ by downfolding the DFT band structure to a bilayer two-orbital model. We find that the leading superconducting instability of La$_3$Ni$_2$O$_7$ has $d_{xy}$ symmetry, if its DFT band structure is accurately reproduced by the downfolded model and the interaction strengths are in a physically reasonable range. More importantly, we show that a slight increase in the crystal field splitting between two Ni-$e_g$ orbitals ($< 0.2$ eV) changes the pairing symmetry from $d_{xy}$ to $s_{\pm}$. We elucidate that such a transition is associated with the change in inverse Fermi velocity and susceptibility, while the shape of Fermi surface remains almost unchanged. Our work highlights the sensitive dependence of La$_3$Ni$_2$O$_7$ pairing symmetry on Ni-$e_g$ crystal field splitting, which provides one source of discrepancy in the pairing symmetry reported in the literature. Such sensitivity calls for care in the procedure of downfolding DFT band structure to low-energy effective models when one calculates the pairing symmetry of multi-orbital superconductors.

We perform DFT calculations, as implemented in the Vienna ab initio simulation package (VASP)~\cite{RevModPhys.64.1045,PhysRevB.54.11169}. We use MLWF to do the downfolding from DFT band structure~\cite{10.3389/fphy.2022.835942} and obtain a bilayer two-orbital model for subsequent linearized gap equation calculations. We refer to our bilayer two-orbital model as Wannier's model. For comparison, we also study another bilayer two-orbital model~\cite{PhysRevLett.131.126001}, referred to as Luo's model in this study, which is widely used to study the superconducting pairing symmetry of La$_3$Ni$_2$O$_7$ in literature. Wannier's model and Luo's model have the same bases: (Ni1-$d_{3z^2-r^2}$, Ni1-$d_{x^2-y^2}$, Ni2-$d_{3z^2-r^2}$, Ni2-$d_{x^2-y^2}$) where 1 and 2 label different NiO$_2$ layers. The difference between the two models is that Wannier's model takes into account all the hoppings between different orbitals, while Luo's model only includes nearest and second-nearest hoppings. Thus the two models have slightly different Ni-$e_g$ crystal field splitting and hopping parameters (see Supplementary Note 3). We use RPA to treat the local Hubbard interaction on Ni-$e_g$ orbitals in the Slater-Kanamori form~\cite{10.1143/PTP.30.275} and obtain the effective pairing potential. The onsite and hopping parameters of Wannier's model and Luo's model, the local Hubbard interaction, as well as the computational details of DFT, RPA and linearized gap equation calculations are found in Methods.

\section{Results}

\subsection{Electronic structure and susceptibility}

\begin{figure}[t]
\includegraphics[angle=0,width=0.9\textwidth]{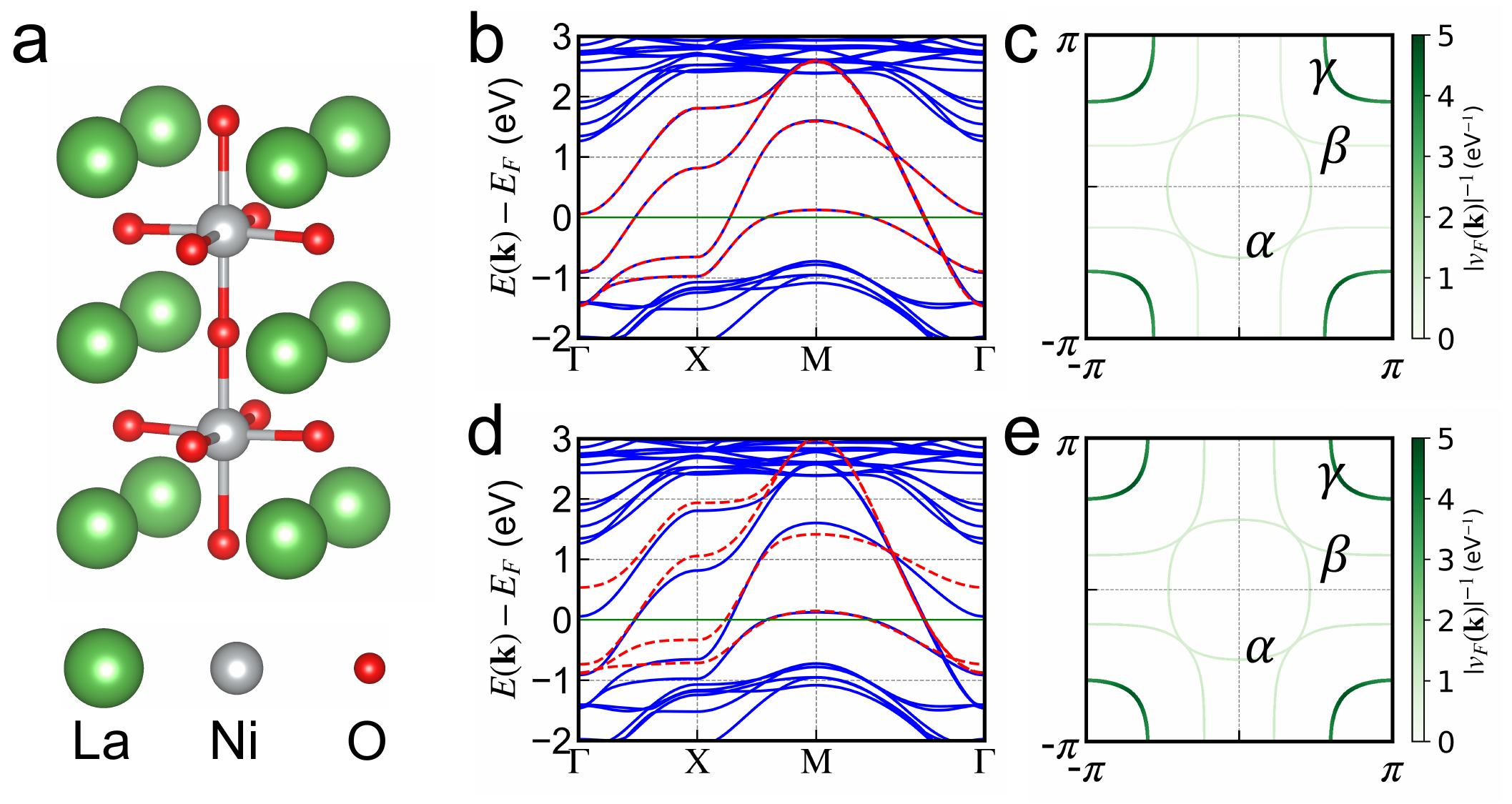}
\caption{\label{fig1} \textbf{Crystal structure and electronic structure.} (a) Crystal structure of La$_3$Ni$_2$O$_7$ under high pressure. (b) Comparison of DFT band structure of La$_3$Ni$_2$O$_7$ under 30 GPa (blue solid curves) to the band structure of Wannier's model (red dashed curves). (c) The Fermi surface of Wannier's model. The green color indicates the magnitude of inverse Fermi velocity $|v_F(\textbf{k})|^{-1}$. (d) Comparison of DFT band structure of La$_3$Ni$_2$O$_7$ under 30 GPa (blue solid curves) to the band structure of Luo's model (red dashed curves). (e) The Fermi surface of Luo's model. The green color indicates the magnitude of inverse Fermi velocity $|v_F(\textbf{k})|^{-1}$. Source data are provided as a Source Data file.}
\end{figure}

Fig.~\ref{fig1}(a) shows the crystal structure of La$_3$Ni$_2$O$_7$ under high pressure. In our DFT calculations, we find that after atomic relaxation, oxygen octahedral rotations in La$_3$Ni$_2$O$_7$ are completely suppressed under 30 GPa, consistent with the experiment~\cite{327nature,wang2023structure,li2024pressuredriven} and other theoretical calculations~\cite{zhang2023structural,geisler2023structural}. We obtain a tetragonal structure whose theoretical lattice constants are in good agreement with the experiment~\cite{327nature,wang2023structure,li2024pressuredriven} (see the comparison in Supplementary Note 1). Following the previous studies~\cite{PhysRevLett.131.126001,yang2023interlayer,liu2023spmwave,PhysRevB.109.104508,PhysRevB.108.L140505,luo2023hightc,zhang2023structural,PhysRevB.108.165141,fan2023superconductivity}, because the inter-block interaction is weak, we focus on one-block of La$_3$Ni$_2$O$_7$, whose DFT band structure is used to do the downfolding to a bilayer two-orbital model whose first Brillouin zone is a simple two-dimensional (2D) square.


Fig.~\ref{fig1}(b) shows the comparison of DFT band structure of La$_3$Ni$_2$O$_7$ and the band structure from Wannier's model. Wannier's model exactly reproduces the four Ni-$e_g$-derived bands close to the Fermi level. The Ni-$e_g$ atomic projections onto those bands are shown in Supplementary Note 4. Fig.~\ref{fig1}(c) shows the Fermi surface of Wannier's model, which has three sheets in the 2D Brillouin zone, labeled as $\alpha$, $\beta$, and $\gamma$. The $\alpha$ and $\beta$ sheets are mainly derived from Ni-$d_{x^2-y^2}$ orbital, while the $\gamma$ sheet dominantly consists of Ni-$d_{3z^2-r^2}$ (see Supplementary Note 4). We use green color to indicate the magnitude of inverse Fermi velocity $|v_F(\textbf{k})|^{-1}$. The $\gamma$ sheet has a large inverse Fermi velocity, while the inverse Fermi velocity of the $\alpha$ and $\beta$ sheets is substantially smaller. Next we switch to Luo's model for comparison. Fig.~\ref{fig1}(d) shows the band structure of Luo's model, which overlays the DFT band structure of La$_3$Ni$_2$O$_7$. Luo's model also reasonably reproduces the four Ni-$e_g$-derived bands around the Fermi level. The minor difference in the band structure between DFT and Luo' model is probably due to the fact that Luo's model only takes into account the nearest and second nearest hoppings~\cite{PhysRevLett.131.126001}, while DFT (as well as Wannier's model) includes all the long-range hoppings. However, the Fermi surface of Luo's model is very similar to that of Wannier's model, as shown in Fig.~\ref{fig1}(e). We find that in Luo's model, the inverse Fermi velocity of the $\gamma$ sheet is slightly larger than that of Wannier's model. We will return to this important point later when we discuss the superconducting pairing symmetry.


\begin{figure}[t]
\includegraphics[angle=0,width=0.9\textwidth]{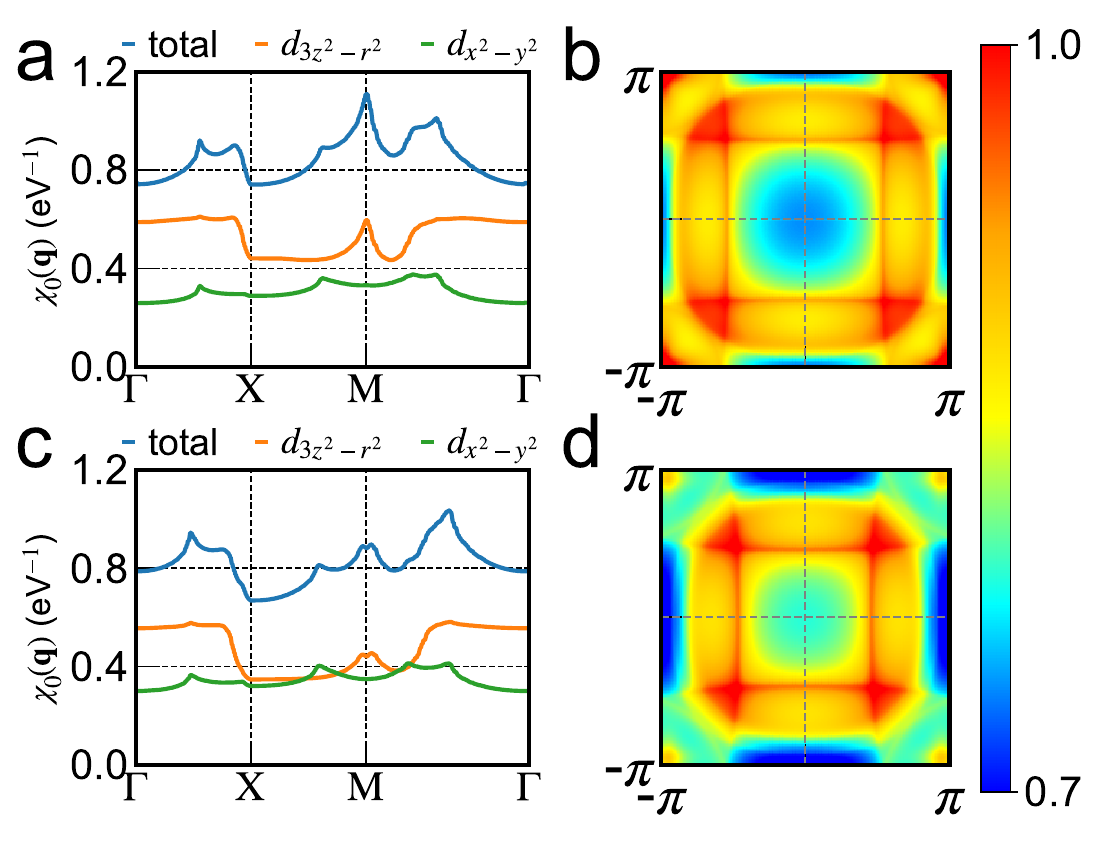}
\caption{\label{fig2} \textbf{Susceptibility.} Bare static
  susceptibility $\chi_0(\textbf{q})$ at $T=116$ K.  (a) Total and
  orbital-projected $\chi_0(\textbf{q})$ of Wannier's model along the
  high-symmetry \textbf{q}-path. (b) Total $\chi_0(\textbf{q})$ of
  Wannier's model in the 2D Brillouin zone.  (c) Total and
  orbital-projected $\chi_0(\textbf{q})$ of Luo's model along the
  high-symmetry \textbf{q}-path. (d) Total $\chi_0(\textbf{q})$ of
  Luo's model in the 2D Brillouin zone. Source data are provided as a
  Source Data file.}
\end{figure}

Next we compare the total and orbital-projected bare susceptibility $\chi_0(\textbf{q})$ of the two models (the definition of different kinds of susceptibility is found in Methods). Fig.~\ref{fig2}(a) and (b) show $\chi_0(\textbf{q})$ of Wannier's model on a high-symmetry $\textbf{q}$-path and in the 2D Brillouin zone. In panel (a), both total and orbital-projected $\chi_0(\textbf{q})$ are shown, while in panel (b), we only show total $\chi_0(\textbf{q})$. In Wannier's model, the total $\chi_0(\textbf{q})$ has a peak at $M$. The peak value would be substantially enhanced in the spin susceptibility $\chi_s(\textbf{q})$ when local Hubbard interactions are added on Ni orbitals in RPA calculations (see Supplementary Note 5). By comparison, we perform the same calculations on Luo's model and show the results in Fig.~\ref{fig2}(c) and (d). We find that while $\chi_0(\textbf{q})$ of Luo's model is overall similar to that of Wannier's model, there are some notable differences. In Luo's model, total $\chi_0(\textbf{q})$ at $M$ gets suppressed and the highest peak of total $\chi_0(\textbf{q})$ is along $\Gamma$ to $M$, close to $\left(\frac{\pi}{2}, \frac{\pi}{2}\right)$. Compared to Wannier's model, the total $\chi_0(\textbf{q})$ is also slightly increased around the zone center ($\Gamma$ point) in Luo's model. The implications of those changes in $\chi_0(\textbf{q})$ for the superconducting pairing symmetry will be discussed at the end.

\subsection{Superconducting pairing symmetry}

\begin{figure}[t]
\includegraphics[angle=0,width=0.9\textwidth]{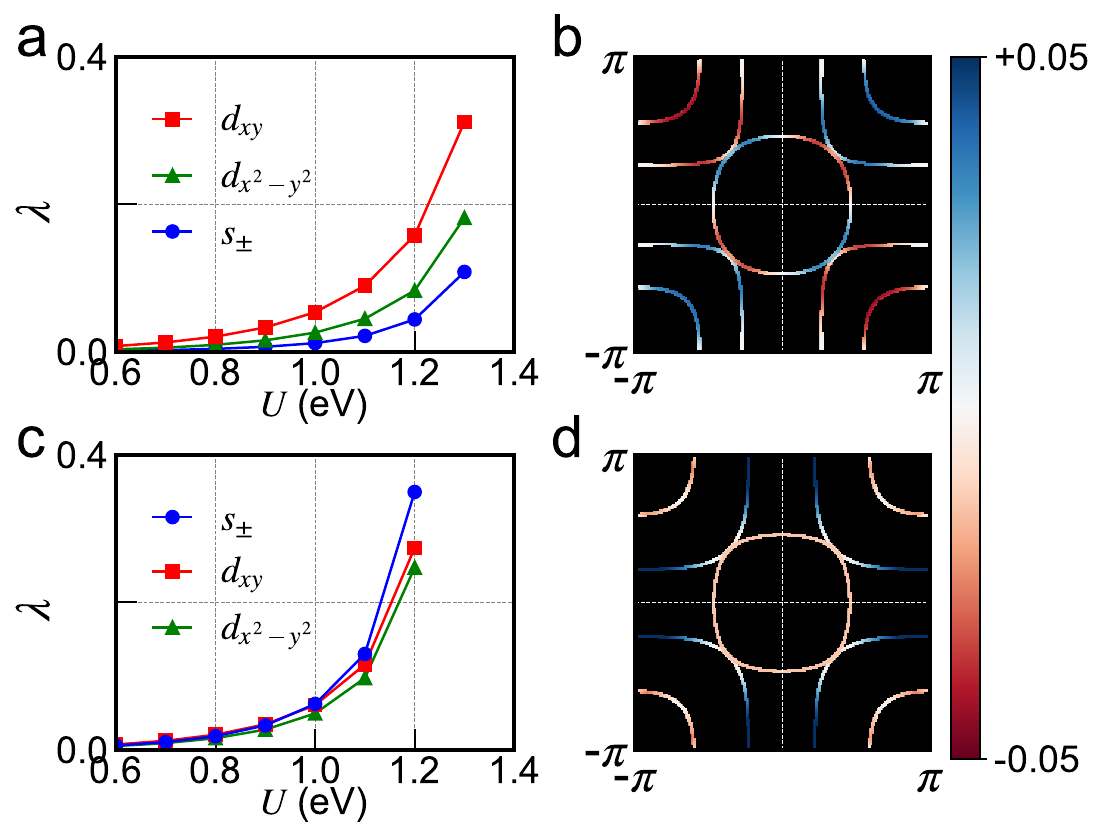}
\caption{\label{fig3} \textbf{Superconducting pairing symmetry.}
(a) The eigenvalues $\lambda$ of linearized gap equation with RPA method, using Wannier's model. The leading superconducting instability has $d_{xy}$ (red squares), while two sub-leading superconducting instabilities have $d_{x^2-y^2}$ (green triangles) and $s_{\pm}$ (blue circles) symmetries, respectively. (b) Gap function for the leading superconducting instability of Wannier's model calculated at $U = 1.2$ eV and $J_{H}=0.15U = 0.18$ eV, which exhibits $d_{xy}$ symmetry.
(c) The eigenvalues $\lambda$ of linearized gap equation with RPA method, using Luo's model. The leading superconducting instability has $s_{\pm}$ (blue circles), while two sub-leading superconducting instabilities have $d_{xy}$ (red squares) and $d_{x^2-y^2}$ (green triangles) symmetries, respectively. (d) Gap function for the leading superconducting instability of Luo's model calculated at $U = 1.2$ eV and $J_{H}=0.15U = 0.18$ eV, which exhibits $s_{\pm}$ symmetry. Source data are provided as a Source Data file.}
\end{figure}

Now we study the superconducting pairing symmetry of La$_3$Ni$_2$O$_7$ by solving the linearized gap equation with RPA method to treat the local Hubbard interaction for the effective pairing potential. RPA method has been widely used to calculate the pairing symmetry of unconventional superconductors such as cuprates and iron-pnictides~\cite{SCALAPINO1995329,annurevconmatphys125055}, which proves to be reliable and consistent with other methods including fluctuation exchange approximation (FLEX) and functional renormalization group (fRG)~\cite{Hirschfeld_2011,RevModPhys.84.1383}. We study a range of $U$ with $J_{H}=0.15U$. The key results do not change if other physically reasonable ratios of $J_{H}/U$ are used (see Supplementary Note 9). In Fig.~\ref{fig3}(a), we show the eigenvalues $\lambda$ of the linearized gap equation using Wannier's model. Since the superconducting transition temperature $T_c \propto e^{-1/\lambda}$, the eigenvector of the largest $\lambda$ corresponds to the most favorable superconducting pairing symmetry. We find that the leading superconducting instability has $d_{xy}$ symmetry with a sign difference between $\alpha$ sheet and $\beta/\gamma$ sheets. In addition, we also show two sub-leading superconducting instabilities that have $d_{x^2-y^2}$ and $s_{\pm}$ pairing symmetries. The leading superconducting pairing symmetry does not change with the interaction strength $U$ (we find $U_c$ = 1.4 eV that triggers the formation of spin density wave). Fig.~\ref{fig3}(b) shows the gap function for the largest eigenvalue of linearized gap equation, using Wannier's model. This gap function clearly exhibits $d_{xy}$ symmetry (the gap function of the two sub-leading pairing symmetries is shown in Supplementary Note 12). In Fig.~\ref{fig3}(c), we show the superconducting pairing symmetry of Luo's model. By contrast, we find that the leading superconducting instability has $s_{\pm}$ symmetry, with two sub-leading superconducting instabilities that have $d_{xy}$ and $d_{x^2-y^2}$ pairing symmetries, consistent with the previous studies~\cite{gu2023effective,zhang2023structural,liu2023spmwave,PhysRevB.108.L140505}. The leading superconducting pairing symmetry in Luo's model does not change with the interaction strength either ($U_c$ = 1.3 eV for spin density wave). Fig.~\ref{fig3}(d) shows the gap function for the largest eigenvalue of linearized gap equation, using Luo's model. This gap function exhibits $s_{\pm}$ symmetry, in which $\alpha$ and $\gamma$ sheets are of the same sign, while $\beta$ sheet is of the opposite sign. We note that a superconducting order parameter that changes sign between different Fermi sheets is widely observed in iron-pnictides~\cite{10.1038/nphys4299,science.1187399,Hirschfeld_2011,RevModPhys.83.1589,PhysRevLett.101.057003,PhysRevB.78.134512}.

\subsection{Sensitivity to Ni-$e_g$ crystal field splitting}

\begin{figure}[t]
\includegraphics[angle=0,width=0.85\textwidth,center]{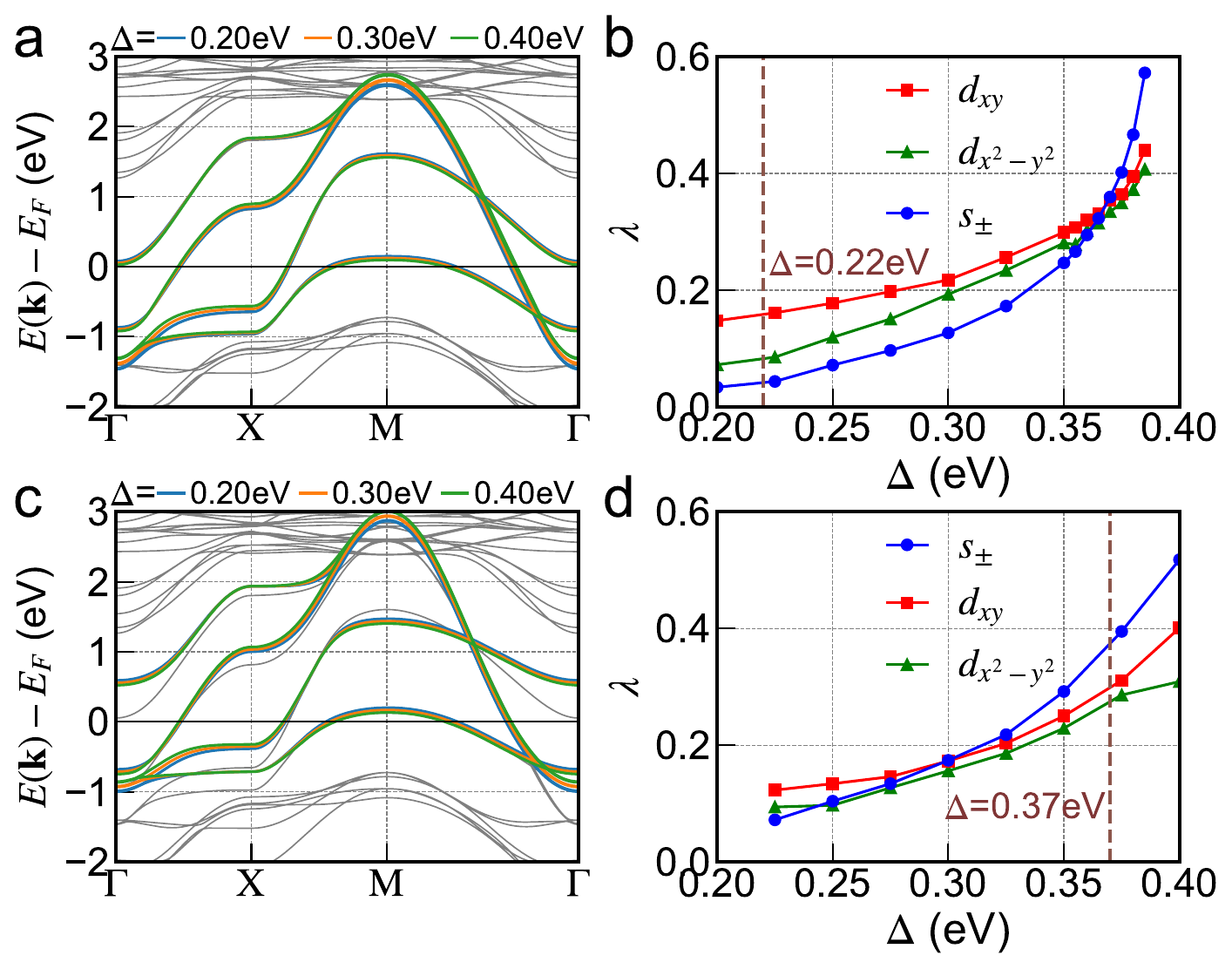}  
\caption{\label{fig4} \textbf{Tuning Ni-$e_g$ crystal field splitting.}
(a) Comparison of DFT band structure of La$_3$Ni$_2$O$_7$ under 30 GPa (gray) to the band structure of modified Wannier's model at a few different Ni-$e_g$ crystal field splittings $\Delta$. (b) The eigenvalues $\lambda$ of linearized gap equation with RPA method for the leading superconducting instability as well as two sub-leading superconducting instabilities in modified Wannier's model, calculated at $U = 1.2$ eV and $J_{H}= 0.18$ eV. The leading superconducting pairing symmetry is $d_{xy}$ ($s_{\pm}$) when $\Delta$ is smaller (larger) than 0.37 eV. The dashed brown line highlights the original value of $\Delta$ = 0.22 eV in Wannier's model. 
(c) Comparison of DFT band structure of La$_3$Ni$_2$O$_7$ under 30 GPa (gray) to the band structure of modified Luo's model at a few different Ni-$e_g$ crystal field splittings $\Delta$. (d) The eigenvalues $\lambda$ of linearized gap equation with RPA method for the leading superconducting instability as well as two sub-leading superconducting instabilities in modified Luo's model, calculated at $U = 1.2$ eV and $J_{H}= 0.18$ eV. The leading superconducting pairing symmetry is $d_{xy}$ ($s_{\pm}$) when $\Delta$ is smaller (larger) than 0.30 eV. The dashed brown line highlights the original value of $\Delta$ = 0.37 eV in Luo's model. Source data are provided as a Source Data file.}
\end{figure}

Next we trace the main source for the different pairing symmetries predicted by Wannier's model and Luo's model, despite the fact that they exhibit very similar Fermi surfaces. As we mentioned earlier, the two models have the same bases but slightly different parameters. By comparing the onsite energies and hopping matrix elements, we find that the ``biggest'' difference between the two models is Ni-$e_g$ crystal field splitting $\Delta$. The difference in $\Delta$ is about 0.15 eV, while the differences in the hopping matrix elements are a few times or even one order of magnitude smaller (see details in Supplementary Note 3). Next we perform a ``thought-experiment'', in which we treat $\Delta$ as a free parameter and manually tune its value. For a given value of $\Delta$, we re-calculate the band structure and superconducting pairing symmetry using the linearized gap equation with RPA method. We use $U=1.2$ eV as a representative interaction strength. Our main results do not depend on this specific value of $U$ (see Supplementary Note 14). In Fig.~\ref{fig4}(a), we compare the band structure of modified Wannier's model to DFT band structure of La$_3$Ni$_2$O$_7$ under 30 GPa, as $\Delta$ is fine-tuned in a small range from 0.20 eV to 0.40 eV. The optimal value of $\Delta$ in Wannier's model is 0.22 eV, based on the fitting. We find that tuning $\Delta$ in this small range yields marginal effects on the overall band structure, since the total bandwidth is about 4 eV. Correspondingly, the shape of the resulting Fermi surfaces remains almost unchanged with $\Delta$ (see Supplementary Note 10). In Fig.~\ref{fig4}(b), we show the eigenvalues for the leading superconducting instability, using modified Wannier's model, as $\Delta$ varies. For comparison, the eigenvalues for two sub-leading superconducting instabilities are also shown. We find that the leading superconducting instability has $d_{xy}$ symmetry ($s_{\pm}$ symmetry) if $\Delta$ is smaller (larger) than 0.37 eV. The brown dashed line highlights the original value of $\Delta$ in Wannier's model, at which the leading superconducting pairing symmetry is $d_{xy}$. Next we switch to Luo's model. In Fig.~\ref{fig4}(c), we compare the band structure of modified Luo's model to DFT band structure of La$_3$Ni$_2$O$_7$ under 30 GPa, as $\Delta$ is fine-tuned in the same range. The optimal value of $\Delta$ in Luo's model is 0.37 eV, larger than Wannier's model. Similar to Wannier's model, we find insignificant changes in the overall band structure of Luo's model by such fine-tuning of $\Delta$. Then we re-calculate the superconducting pairing symmetry at the same interaction strengths, using modified Luo's model, as $\Delta$ varies. Fig.~\ref{fig4}(d) shows the eigenvalues for the leading superconducting instabilities, as well as two sub-leading superconducting instabilities. We find the same trend that a smaller (larger) $\Delta$ favors $d_{xy}$ symmetry ($s_{\pm}$ symmetry) with the critical point at 0.30 eV. The dashed brown line highlights the original value of $\Delta$ in Luo's model, which is larger than the critical value. Thus the leading superconducting pairing symmetry of Luo's model is $s_{\pm}$. 

\begin{figure}[t]
\includegraphics[angle=0,width=0.85\textwidth,center]{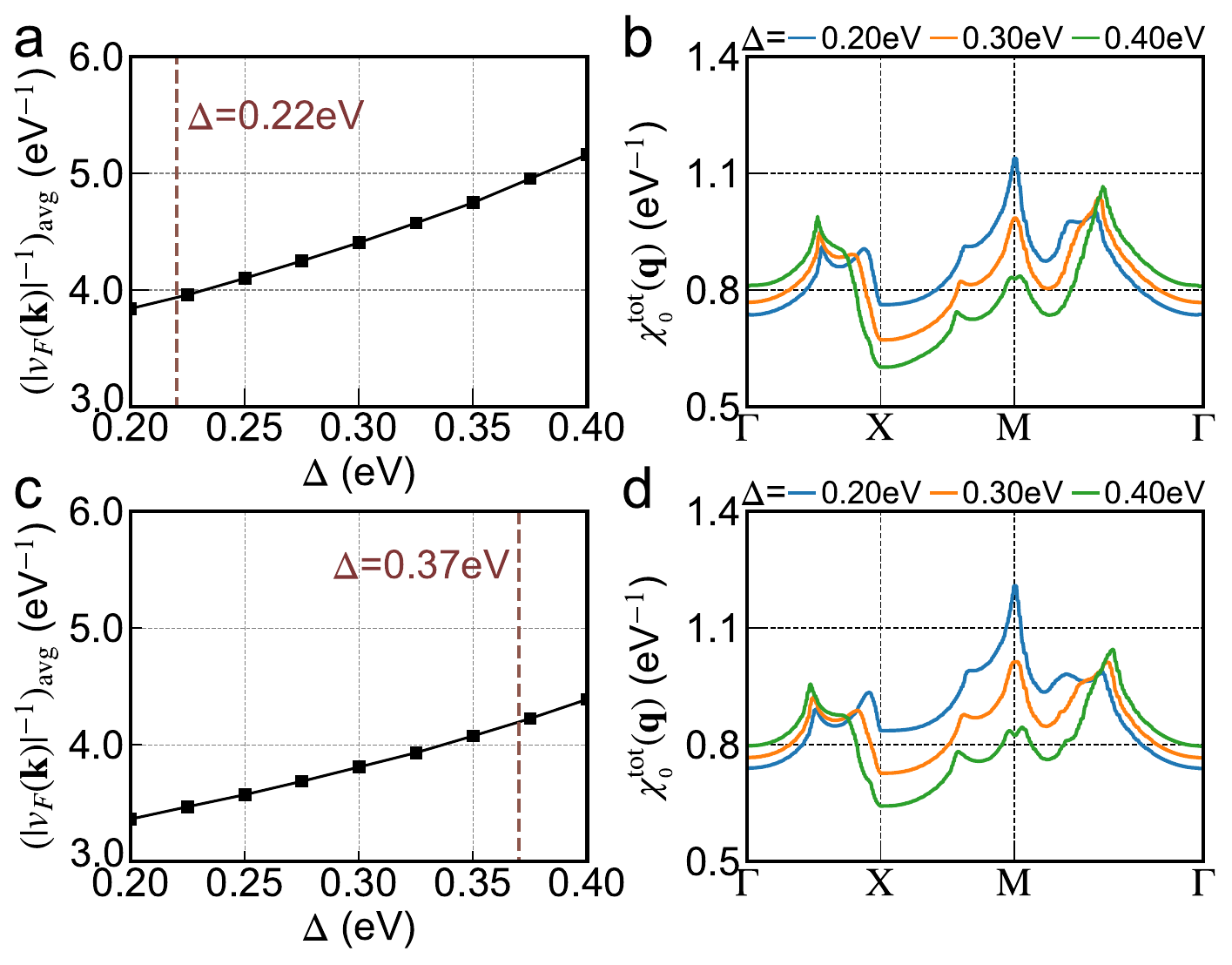}  
\caption{\label{fig5} \textbf{Inverse Fermi velocity and bare static susceptibility.} (a) Average inverse Fermi velocity $(|v_F(\textbf{k})|^{-1})_{\rm{avg}}$ over the $\gamma$ sheet of modified Wannier’s model as a function of Ni-$e_g$ crystal field splitting $\Delta$. The dashed brown line highlights the original value of $\Delta$ = 0.22 eV in Wannier's model. (b) Total $\chi_0(\textbf{q})$ along the high-symmetry \textbf{q}-path of modified Wannier's model at a few different Ni-$e_g$ crystal field splittings $\Delta$.
(c) Average inverse Fermi velocity $(|v_F(\textbf{k})|^{-1})_{\rm{avg}}$ over the $\gamma$ sheet of modified Luo’s model as a function of Ni-$e_g$ crystal field splitting $\Delta$. The dashed brown line highlights the original value of $\Delta$ = 0.37 eV in Luo's model. (d) Total $\chi_0(\textbf{q})$ along the high-symmetry \textbf{q}-path of modified Luo's model at a few different Ni-$e_g$ crystal field splittings $\Delta$.  We set the lattice constant $a=1$ and the reduced Planck constant $\hbar = 1$, and thus Fermi velocity has the unit of energy. Source data are provided as a Source Data file.}
\end{figure}

To get a deeper understanding of the effects from a small change in Ni-$e_g$ crystal field splitting on the pairing symmetry, we study the inverse Fermi velocity $|v_F(\textbf{k})|^{-1}$ and bare total susceptibility $\chi_0(\textbf{q})$, both of which play a crucial role in the linearized gap equation calculations~\cite{Graser_2009}. As we find in Fig.~\ref{fig1}, $|v_F(\textbf{k})|^{-1}$ has large values on $\gamma$ sheet, while the $\alpha$ and $\beta$ sheets have substantially smaller $|v_F(\textbf{k})|^{-1}$. Thus we average $|v_F(\textbf{k})|^{-1}$ over the $\gamma$ sheet of modified Wannier's model and show $(|v_F(\textbf{k})|^{-1})_{\rm{avg}}$ as a function of Ni-$e_g$ crystal field splitting $\Delta$. The raw data of $|v_F(\textbf{k})|^{-1}$ as a function of $\Delta$ are found in Supplementary Note 10. Fig.~\ref{fig5}(a) shows that $(|v_F(\textbf{k})|^{-1})_{\rm{avg}}$ monotonically increases with $\Delta$. This indicates that with an increased $\Delta$, the $\gamma$ sheet that is mainly composed of $d_{3z^2-r^2}$ orbital, carries more weight in the linearized gap equation calculations. Fig.~\ref{fig5}(b) shows the bare total susceptibility $\chi_0(\textbf{q})$ of modified Wannier's model at a few different $\Delta$ (the orbital-projected $\chi_0(\textbf{q})$ are shown in Supplementary Note 11). Fig.~\ref{fig5}(b) shows that when $\Delta = 0.2$ eV, total $\chi_0(\textbf{q})$ has a clear peak at $M$ point, which suggests that intralayer correlations are significant. This usually favors $d$-wave pairing~\cite{RevModPhys.72.969,keimer2015quantum,PhysRevLett.110.216405}. As $\Delta$ increases, total $\chi_0(\textbf{q})$ gets reduced along the zone boundary ($X \to M$), but increases around the zone center ($\Gamma$ point). The enhancement of $\chi_0(\textbf{q})$ around $\Gamma$ point is associated with the strengthening of interlayer correlations. This, together with $d_{3z^2-r^2}$ orbital carrying more weight, favors $s$-wave pairing, as a number of recent studies have shown~\cite{PhysRevB.108.L140504,PhysRevB.108.174511,PhysRevLett.132.146002}. Fig.~\ref{fig5}(c) and (d) show $(|v_F(\textbf{k})|^{-1})_{\rm{avg}}$ and total $\chi_0(\textbf{q})$ of modified Luo's model as $\Delta$ varies. We find exactly the same trends as Wannier's model: 1) $(|v_F(\textbf{k})|^{-1})_{\rm{avg}}$ increases with $\Delta$ and 2) an increasing $\Delta$ reduces $\chi_0(\textbf{q})$ along the zone boundary but enhances $\chi_0(\textbf{q})$ around the zone center, both of which suggests that the favorable superconducting pairing symmetry would change from $d$-wave to $s$-wave. In particular, at the original value of $\Delta$, $(|v_F(\textbf{k})|^{-1})_{\rm{avg}}$ of Luo's model is 4.2 eV$^{-1}$, larger than that of Wannier's model (3.9 eV$^{-1}$), which shows the tendency of favoring $s_{\pm}$ symmetry.

\section{Discussions}

Before we conclude, we make a few comments.

1) The above analysis does not mean that the pairing symmetry of La$_3$Ni$_2$O$_7$ only depends on Ni-$e_g$ crystal field splitting. Rather, what we want to highlight is the sensitive dependence of La$_3$Ni$_2$O$_7$ pairing symmetry on its low-energy electronic structure, in particular Ni-$e_g$ crystal field splitting. This implies that a prediction of pairing symmetry based on a given tight-binding model could be biased and does not provide the complete picture, considering inevitable approximations and uncertainties introduced in the modelling. Instead, the trend that a smaller (larger) Ni-$e_g$ crystal field splitting favors $d_{xy}$ ($s_{\pm}$) pairing symmetry in La$_3$Ni$_2$O$_7$ is robust and conclusive.


2) Compared to Ref.~\cite{christiansson2023correlated}, we employ smaller $U$ values in the linearized gap equation calculations with RPA, because RPA is a weak-coupling theory and $U$ in RPA calculations is understood as a renormalized (i.e. reduced) interaction $\bar{U}$~\cite{PhysRevB.47.2742,PhysRevLett.66.369}. Spin density wave (SDW) instability is triggered at a small $\bar{U}$ value in the RPA method. This is an intrinsic property of RPA, as pointed out by Refs.~\cite{Graser_2009,PhysRevLett.66.369}. Our critical $U$ value for SDW in La$_3$Ni$_2$O$_7$ is consistent with the previous RPA studies~\cite{liu2023spmwave,PhysRevB.108.L201121}. 

3) Our calculations show that when the ratio $J_H/U$ is within a physically reasonable range between 0.1 and 0.2~\cite{PhysRevB.108.L140505,gu2023effective,PhysRevB.108.L201121,PhysRevB.101.060504,liu2023spmwave}, Wannier's model favors $d_{xy}$ symmetry. Using Wannier's model, we also calculate the phase diagram of pairing symmetry for a larger range of $J_H/U$ ratios (see Supplementary Note 9). We find that when $J_H/U$ were sufficiently large, Wannier's model would favor $d_{x^2-y^2}$ pairing symmetry, consistent with Ref.~\cite{PhysRevB.108.L201121}. In addition, a number of other works use variants of $t$-$J$ model to study the superconducting pairing symmetry of La$_3$Ni$_2$O$_7$~\cite{PhysRevLett.132.146002,PhysRevB.108.174511,PhysRevB.108.214522}. They find that the pairing symmetry depends on exchange interactions $J$. This does not conflict with our results, because in multi-orbital systems, when one derives the $t$-$J$ model from a Hubbard-like model, the exchange interaction $J$ depends on hopping parameters and crystal field splitting, as well as $U$ and $J_H$. In our work, we find that when Ni-$e_g$ crystal field splitting increases, the favorable pairing symmetry changes from $d$-wave to $s$-wave, because the $d_{3z^2-r^2}$ orbital carries more weight in the linearized gap equation calculations and the interlayer correlation is enhanced, as manifested by the susceptibility $\chi(\textbf{q})$ at $\Gamma$ point. This is consistent with Refs.~\cite{PhysRevLett.132.146002,PhysRevB.108.174511}, which find that an increased interlayer exchange interaction $J_{\perp}$ would change the pairing symmetry from $d$-wave to $s$-wave in La$_3$Ni$_2$O$_7$.



4) We comment that tuning Ni-$e_g$ crystal field splitting of complex oxides is experimentally feasible because Ni-$e_g$ crystal field splitting is directly coupled to the NiO$_6$ oxygen octahedra environment~\cite{PhysRevB.102.155148,HWANG2019100}. Static or dynamical perturbations may change the Ni-$e_g$ crystal field splitting by up to 0.5 eV, which is sufficiently large to alter the pairing symmetry of La$_3$Ni$_2$O$_7$.

5) Related to the preceding point, since the physical Ni-$e_g$ crystal field splitting is very close to the critical value, La$_3$Ni$_2$O$_7$ may exhibit distinct pairing symmetries under experimental perturbations that can either alter the ratio of in-plane and out-of-plane Ni-O bond lengths, or move the position of Ni atoms in a NiO$_6$ oxygen octahedron.


In conclusion, by performing first-principles and effective model calculations, we find that the superconducting pairing symmetry of La$_3$Ni$_2$O$_7$ very sensitively depends on Ni-$e_g$ crystal field splitting. A slight increase in Ni-$e_g$ crystal field splitting ($< 0.2$ eV) changes the leading superconducting pairing symmetry from $d_{xy}$ to $s_{\pm}$. When the DFT band structure is exactly reproduced by the downfolded model, we find that the pairing symmetry of La$_3$Ni$_2$O$_7$ is $d_{xy}$. We also reveal that the transition from $d_{xy}$ to $s_{\pm}$ pairing symmetry is associated with an increase in inverse Fermi velocity and zone-center susceptibility. Finally we conjecture that such sensitive dependence of pairing symmetry on low-energy electronic structure may exist in other multi-orbital unconventional superconductors (e.g. La$_4$Ni$_3$O$_{10}$~\cite{PhysRevB.109.144511,li2023signature,li2024structural,zhang2024spmwave,luo2024trilayer,zhang2024prediction,zhu2024superconductivity,zhang2024superconductivity}) and thus care is required in the downfolding procedure when ones studies unconventional superconductivity in realistic materials.




\section*{Methods}

\def\p{\partial}
\def\a{\alpha}
\def\b{\beta}
\def\lg{\left\langle}
\def\rg{\right\rangle}
\def\d{\mathrm{d}}
\def\f{\mathbf{f}}
\def\g{\mathbf{g}}
\def\x{\mathbf{x}}
\def\w{\wedge}
\def\q{\mathbf{q}}
\def\k{\mathbf{k}}
\def\n{\nabla}
\def\e{\mathrm{e}}
\def\ep{\mathrm{\epsilon}}
\def\ra{\rightarrow}
\def\R{\mathbf{R}}
\def\G{\mathbf{G}}
\def\ll{\left|}
\def\rr{\right|}
\def\[{\begin{equation}\begin{aligned}}
\def\]{\end{aligned}\end{equation}}
\def\beq{\begin{equation}}
\def\eeq{\end{equation}}
\def\bma{\begin{bmatrix}}
\def\ema{\end{bmatrix}}
\def\kg{\text{ }}

\subsection{First-Principles Calculations}

We perform DFT~\cite{PhysRev.136.B864,PhysRev.140.A1133} calculations, as implemented in VASP~\cite{RevModPhys.64.1045,PhysRevB.54.11169} with the projector augmented wave (PAW) method. We use generalized gradient approximation (GGA) with Perdew-Burke-Ernzerhof parametrization~\cite{PhysRevLett.77.3865} for the exchange-correlation functional. We use an energy cutoff of 600 eV. We employ the conventional orthogonal cell of La$_3$Ni$_2$O$_7$ (48-atom) to perform the atomic relaxation. The $\textbf{k}$-mesh is $16\times 16\times 4$. The total energy convergence criterion is 10$^{-7}$ eV. The force convergence criterion is 1 meV/\AA. The pressure convergence criterion is 0.1kbar. We find that under a pressure of 30 GPa, oxygen octahedra rotations in La$_3$Ni$_2$O$_7$ are completely suppressed and La$_3$Ni$_2$O$_7$ is stabilized in a tetragonal structure (space group $I4/mmm$, see Supplementary Note 13 for details). The DFT optimized lattice constants and internal atomic positions, as well as the experimental lattice constants are shown in Supplementary Note 1.




We note that the bilayer two-orbital model is intended to study a monoblock of La$_3$Ni$_2$O$_7$ because the inter-block interaction is weak~\cite{PhysRevLett.131.126001,PhysRevB.108.L140505,PhysRevB.108.174501,PhysRevB.108.165141,luo2023hightc,gu2023effective,zhang2023structural}. Thus we focus on a monoblock of La$_3$Ni$_2$O$_7$ with its optimized atomic positions fixed and a vacuum of 30~\AA~thick inserted in the cell. The corresponding Brillouin zone is a two-dimensional simple square. We perform a self-consistent calculation with a $\textbf{k}$-mesh of $20\times 20 \times 1$ on this bulk-like ``freestanding'' La$_3$Ni$_2$O$_7$ to make direct connections to the bilayer two-orbital model in literature~\cite{PhysRevLett.131.126001,PhysRevB.108.L140505,PhysRevB.108.174501,PhysRevB.108.165141,luo2023hightc,gu2023effective,zhang2023structural} (see Fig.~\ref{fig1}(a) in the main text for the crystal structure of La$_3$Ni$_2$O$_7$ under high pressure).

\subsection{The Downfolded Models}
The details of Wannier's model and Luo's model are shown below. We set the in-plane lattice constants $a=b$ to be 1 throughout. Both models can be compactly written as:
\[\label{eq:0}
\hat{H} = \hat{H}_0 + \hat{H}_{\rm{int}} 
\]
where
\[\label{eq:01}
\hat{H}_0 = \sum_{\textbf{k}\sigma} \hat{\Psi}_{\textbf{k}\sigma}^{\dagger}\mathcal{H}_0(\textbf{k})\hat{\Psi}_{\textbf{k}\sigma}
\]
and $\hat{\Psi}_{\textbf{k}\sigma}^{\dagger} = (\hat{c}^{\dagger}_{\k\sigma 1z}, \hat{c}^{\dagger}_{\k\sigma 1x}, \hat{c}^{\dagger}_{\k\sigma 2z}, \hat{c}^{\dagger}_{\k\sigma 2x})$ in which $\hat{c}_{\k\sigma \nu}^{\dagger}$ is the creation operator for orbital $\nu$ with momentum $\k$ and spin $\sigma$. $(1z, 1x, 2z, 2x)$ is a shorthand for the orbital basis:
\[\label{eq:1}
|\textrm{Ni1-}d_{3z^2-r^2}\rangle, |\textrm{Ni1-}d_{x^2-y^2}\rangle,|\textrm{Ni2-}d_{3z^2-r^2}\rangle,|\textrm{Ni2-}d_{x^2-y^2}\rangle
\]
where 1 and 2 label different NiO$_2$ layers.

For $\hat{H}_{\rm{int}}$, we use the full Slater-Kanomori-type interaction for multi-orbital systems~\cite{10.1143/PTP.30.275}:
\begin{eqnarray}
  \label{eqSr}  \nonumber \hat{H}_{\rm{int}} = U\sum_{\alpha,\nu}\hat{n}_{\alpha \nu\uparrow}\hat{n}_{\alpha \nu \downarrow} + \sum_{\alpha, \sigma\sigma'}\Big[(U'-J_H\delta_{\sigma \sigma'}) (\hat{n}_{\alpha 1z \sigma} \hat{n}_{\alpha 1x \sigma'} +  \hat{n}_{\alpha 2z \sigma} \hat{n}_{\alpha 2x \sigma'})\Big] \\ \nonumber
  -J_{H}\sum_{\alpha} \left(  \hat{c}^{\dagger}_{\alpha 1z \downarrow}\hat{c}^{\dagger}_{\alpha 1x \uparrow}\hat{c}_{\alpha 1x \downarrow}\hat{c}_{\alpha 1z \uparrow} + \hat{c}^{\dagger}_{\alpha 2z \downarrow}\hat{c}^{\dagger}_{\alpha 2x \uparrow}\hat{c}_{\alpha 2x \downarrow}\hat{c}_{\alpha 2z \uparrow} + \textrm{h.c.} \right) \\ 
-J'\sum_{\alpha} \left(  \hat{c}^{\dagger}_{\alpha 1z \downarrow}\hat{c}^{\dagger}_{\alpha 1z \uparrow}\hat{c}_{\alpha 1x \downarrow}\hat{c}_{\alpha 1x \uparrow} + \hat{c}^{\dagger}_{\alpha 2z \downarrow}\hat{c}^{\dagger}_{\alpha 2z \uparrow}\hat{c}_{\alpha 2x \downarrow}\hat{c}_{\alpha 2x \uparrow} + \textrm{h.c.} \right) 
\end{eqnarray}  
where $\alpha$ labels a site, $\nu$ labels an atomic orbital ($\nu = 1x, 1z, 2x, 2z$) and $\sigma$ is spin. `h.c.' means hermitian conjugate. $U$ and $U'$ are the intraorbital and interorbital interaction strengths. $J_{H}$ is the Hund's coupling strength and $J'$ is the pair-hopping strength. The first line of Eq.~(\ref{eqSr}) is the density-density interactions. The second line is the spin-flip interaction. The third line is the pair-hopping interaction. Throughout our calculations, we set $U'=U-2J_{H}$ and $J'=J_{H}$.

Wannier's model is obtained by using MLWF~\cite{RevModPhys.84.1419,MOSTOFI2008685} to fit the DFT band structure of La$_3$Ni$_2$O$_7$ under 30 GPa. $\mathcal{H}^{\rm{W}}_0(\textbf{k})$ is written in the form of:
\[\label{eq:5}
\mathcal{H}^{\rm{W}}_0(\k) = \sum_{\mathbf{R}}\mathcal{H}^{\rm{W}}_0(\mathbf{R})e^{i\k\cdot \mathbf{R}}
\]
We use a $\Gamma$-centered \textbf{k}-mesh of $20\times 20$ to sample the two-dimensional Brillouin zone. Hence there are altogether 441 $\textbf{R}$ and $|R_x|\le 10$ and $|R_y|\le 10$. The onsite and hopping matrix elements up to the third-nearest-neighbor $\mathcal{H}^{\rm{W}}_0(\mathbf{R})$ are given below (in the unit of eV):
\[\label{eq:6}
\mathcal{H}^{\rm{W}}_0(0,0) = \bma
+1.153& +0.000& -0.602& +0.000\\
+0.000& +1.375& +0.000& +0.010\\
-0.602& +0.000& +1.153& +0.000\\
+0.000& +0.010& +0.000& +1.375\ema
\]\[
\mathcal{H}^{\rm{W}}_0(1,0) = \bma
-0.155& -0.248& +0.030& +0.031\\ 
-0.248& -0.487& +0.031& +0.000\\ 
+0.030& +0.031& -0.155& -0.248\\ 
+0.031& +0.000& -0.248& -0.487\ema
\]
\[\label{eq:7}
\mathcal{H}^{\rm{W}}_0(1,1) =\bma
-0.023& +0.000& +0.004 & +0.000\\ 
+0.000& +0.064& +0.000 & +0.003\\ 
+0.004& +0.000& -0.023 & +0.000\\ 
+0.000& +0.003& +0.000 & +0.064\ema
\]\[
\mathcal{H}^{\rm{W}}_0(2,0) = \bma
 -0.011& +0.022& -0.003& +0.003\\ 
 +0.022& -0.064& +0.003& -0.004\\ 
 -0.003& +0.003& -0.011& +0.022\\ 
 +0.003& -0.004& +0.022& -0.064\ema
  \]
The $C_4$ rotation symmetry in the tetragonal cell leads to :
\[\label{eq:8}
&\mathcal{H}^{\rm{W}}_0(1,0) = \mathcal{H}^{\rm{W}}_0(-1,0) = \mathcal{H}^{\rm{W}}_0(0,1)=\mathcal{H}^{\rm{W}}_0(0,-1)\\ 
&\mathcal{H}^{\rm{W}}_0(1,1) = \mathcal{H}^{\rm{W}}_0(1,-1) = \mathcal{H}^{\rm{W}}_0(-1,1) = \mathcal{H}^{\rm{W}}_0(-1,-1)\\
&\mathcal{H}^{\rm{W}}_0(2,0) = \mathcal{H}^{\rm{W}}_0(-2,0) = \mathcal{H}^{\rm{W}}_0(0,2)=\mathcal{H}^{\rm{W}}_0(0,-2)
\]
Other $\mathcal{H}_0(\textbf{R})$ can be found in our Supplementary File ``wannier90\_hr.dat''. Using the occupancy of 3 electrons per formula, we obtain that the Fermi level of Wannier's model is 0.782 eV. We note that for the furthest distance $\textbf{R}=(10,10)$, the magnitude of each matrix element of $\mathcal{H}^{\rm{W}}_0(10,10)$ does not exceed 0.03 meV. This shows that Wannier's model practically includes all nonzero hoppings.

Luo's model is constructed in Ref.~\cite{PhysRevLett.131.126001}, which takes the form of:
\[\label{eq:2}
\mathcal{H}^{\rm{L}}_0(\k) = \bma
\mathcal{H}_{11}(\k)& \mathcal{H}_{12}(\k)\\
\mathcal{H}_{12}(\k)& \mathcal{H}_{22}(\k)
\ema
\]
where
\[\label{eq:3}
\mathcal{H}_{11}(\k)= \mathcal{H}_{22}(\k)=\bma
T_{\k}^z & V_{\k} \\
V_{\k} & T_{\k}^x
\ema, \quad \mathcal{H}_{12}(\k)=\bma
t_{\perp}^z & V_{\k}^{\prime} \\
V_{\k}^{\prime} & t_{\perp}^x
\ema
\]
with
\[\label{eq:4}
\begin{aligned}
T_{\k}^{x / z} & =2 t_1^{x / z}\left(\cos k_x+\cos k_y\right)+4 t_2^{x / z} \cos k_x \cos k_y+\epsilon^{x / z} \\
V_{\k} & =2 t_3^{x z}\left(\cos k_x-\cos k_y\right), \quad V_{\k}^{\prime}=2 t_4^{x z}\left(\cos k_x-\cos k_y\right).
\end{aligned}
\]
All the parameters in Luo's model are found in Table~\ref{tab:1}.

\begin{table*}[h!]
  \caption{The parameters of Luo's model $\mathcal{H}_0^{\rm{L}}(\k)$~\cite{PhysRevLett.131.126001}. The unit is eV.}
\begin{ruledtabular}
  \begin{tabular}{ccccc}
    $ t_1^x$ & $t_1^z$ & $t_2^x$ & $t_2^z$ & $t_3^{x z}$ \\\colrule
    -0.483 & -0.110 & 0.069 & -0.017 & 0.239\\\colrule
    $t_{\perp}^x$ & $t_{\perp}^z$ & $t_4^{x z}$ & $\epsilon^x$ & $\epsilon^z$ \\\colrule
    0.005 & -0.635 & -0.034 & 0.776 & 0.409 
    \end{tabular}
\end{ruledtabular}
\label{tab:1}
\end{table*}

\subsection{Susceptibility}

We define different types of susceptibility as follow. We set the Boltzmann constant $k_B$ and the reduced Planck constant $\hbar$ to be 1 throughout. For a multi-orbital non-interacting model $\mathcal{H}_0(\textbf{k})$, the bare susceptibility is defined as~\cite{PhysRevB.101.060504}:
\[\label{eq:11}
\chi_0(q)_{l_1l_2,l_3l_4}&=-\sum_k [G_0(k+q)]_{l_1l_3}[G_0(k)]_{l_4l_2}\\
\]
where $q=(\q, \Omega_m)$ and $k=(\k, \omega_n)$ are four-component vectors. $\Omega_m = 2m\pi T$ is a bosonic frequency and $\omega_n = (2n+1)\pi T$ is a fermionic frequency. $T$ is the temperature and the summation $\sum_k=\frac{T}{N_\k}\sum_{\k,\omega_n}$. For non-interacting Green function $G_0$, the summation over $\omega_n$ can be analytically performed and we obtain:
\[\label{eq:11aa}
\chi_0(q)_{l_1l_2,l_3l_4}= -\frac{1}{N_\k}\sum_\k\sum_{n,n'}(a_{\k+\q})_{l_1 n}(a_{\k+\q})^{\ast}_{l_3 n }
	(a_{\k})_{l_4 n'}(a_{\k})^{\ast}_{ l_2 n'}\frac{n_F[(\ep_{\k+\mathbf{q}})_{n}]-n_F[(\ep_\k)_{n'}]}{(\ep_{\k+\mathbf{q}})_{n}-i\Omega_m-(\ep_\k)_{n'}}
\]
where $a_\k$ is the transition matrix that diagonalizes $\mathcal{H}_0(\k)$. $n_F$ is the Fermi-Dirac occupancy function and $\ep({\k})_n$ is the $n$-th eigenvalue of $\mathcal{H}_0(\k)$. If $\Omega_m = 0$, we get the static susceptibility $\chi_0(\q)=\chi_0(\q,\Omega_m=0)$:
\[\label{eq:11a}
\chi_0(\textbf{q})_{l_1l_2,l_3l_4}=-\frac{1}{N_\k}\sum_\k\sum_{n,n'}(a_{\k+\q})_{l_1 n}(a_{\k+\q})^{\ast}_{l_3 n }
	(a_{\k})_{l_4 n'}(a_{\k})^{\ast}_{ l_2 n'}\frac{n_F[(\ep_{\k+\mathbf{q}})_{n}]-n_F[(\ep_\k)_{n'}]}{(\ep_{\k+\mathbf{q}})_{n}-(\ep_\k)_{n'}}
\]

The above bare static susceptibility is a four-index tensor. The total bare static susceptibility is defined as:
\[\label{eq:11b}
\chi_0^{\rm{tot}}(\textbf{q}) = \frac{1}{2}\sum_{l,l'}[\chi_0(\textbf{q})]_{ll,l'l'}
\]
The orbital projected bare static susceptibility is defined as:
\[\label{eq:11bb}
\chi_0^{\nu}(\textbf{q}) = \frac{1}{2}\sum_{l,l' = \nu}[\chi_0(\textbf{q})]_{ll,l'l'}
\]
where $\nu$ is a given orbital. In the bilayer two-orbital model, $\nu$ is either $d_{3z^2-r^2}$ or $d_{x^2-y^2}$.

The static spin and charge susceptibilities within random-phase-approximation are defined in the next subsection.

\subsection{\label{sec:RPA}Random Phase Approximation Calculations}

The details of random-phase-approximation (RPA)~\cite{PhysRevB.31.4403,YANASE20031,Graser_2009,RevModPhys.84.1383,PhysRevLett.123.247001,PhysRevB.101.060504} method for the calulations of spin and charge susceptibilities are shown below. Based on charge and spin susceptibilities, we construct an effective pairing potential for subsequent linearized gap equation calculations.

Throughout this section, the tensor multiplication of two four-index tensors $C=AB$ is defined as follows:
\[\label{eq:9}C_{ij,kl} = (AB)_{ij,kl} = \sum_{m,n}A_{ij,mn}B_{mn,kl}.
\]
where both $A$ and $B$ are four-index tensors. The spin and charge susceptibilities within RPA method are defined as~\cite{YANASE20031,Graser_2009,PhysRevLett.123.247001}:
\[\label{eq:12}
\chi_{s}(\q) &= [I- \chi_0(\q)U_s]^{-1}\chi_{0}(\q)
\]
\[
\chi_{c}(\q) &= [I+ \chi_0(\q)U_c]^{-1}\chi_{0}(\q)
\]
where the $U$-matrices take the form of~\cite{PhysRevB.101.060504}:
\[\label{eq:14}
\begin{aligned}
(U_s)_{\alpha l_1, \alpha l_2, \alpha l_3, \alpha l_4}&= \begin{cases}U_{\alpha} & l_1=l_2=l_3=l_4 \\
U_\alpha^{\prime} & l_1=l_3 \neq l_2=l_4 \\
J_{H\alpha} & l_1=l_2 \neq l_3=l_4 \\
J_\alpha^{\prime} & l_1=l_4 \neq l_2=l_3\end{cases} \\
(U_c)_{\alpha l_1, \alpha l_2 ; \alpha l_3, \alpha l_4}&= \begin{cases}U_\alpha & l_1=l_2=l_3=l_4 \\
-U_\alpha^{\prime}+2 J_{H\alpha} & l_1=l_3 \neq l_2=l_4 \\
2 U_\alpha^{\prime}-J_{H\alpha} & l_1=l_2 \neq l_3=l_4 \\
J_\alpha^{\prime} & l_1=l_4 \neq l_2=l_3\end{cases}
\end{aligned}
\]
where $\alpha$ labels a site and $l_i$ $(i=1,2,3,4)$ labels an orbital.
All the other entries of $U_c$, $U_s$ are zero. The derivation of Eq.~(\ref{eq:14}) based on the Hamiltonian Eq.~(\ref{eqSr}) is found Supplementay Note 4.

In this work, we choose $U'_{\alpha} = U_{\alpha} -2J_{H\alpha}$ and $J_{\alpha}'=J_{H\alpha}$. For the results in the main text, we use $J_{H}=0.15U$ for each Ni site~\cite{PhysRevB.101.060504}. We also test other $J_{H}/U$ ratios and find that the key results do not qualitatively change (see Supplementary Note 9). 

The effective pairing potential in singlet (s) or triplet (t) channel is given by~\cite{Graser_2009}
\[\label{eq:15}
V^{\rm{s/t}}(\k,\k') &= \frac{1}{2}[V(\k,\k') \pm V(\k,-\k')]
\]
where
\[\label{eq:16}
V_{ij, mn}\left(\k, \k^{\prime}\right)=\sum_{l_1l_2l_3l_4} (a_{\k})^*_{l_2 m}(a_{-\k})^*_{l_3 n}\operatorname{Re}\left[\mathcal{V}_{l_1l_2,l_3l_4}(\k,\k')\right](a_{\k^{\prime}})_{l_1 i}  (a_{-\k^{\prime}})_{l_4 j}  
\]
with
\[\label{eq:17}
\mathcal{V}_{l_1l_2, l_3l_4}(\k,\k')=
\left\{\begin{aligned}
    \frac{1}{2}\left[U_s+U_c+3 {U_s} \chi_s\left(\k-\k'\right) U_s-{U_c} \chi_c \left(\k-\k'\right) U_c\right]_{l_1l_2,l_3l_4} \quad &\text{for singlet}\\
    -\frac{1}{2}\left[ {U_s} \chi_s\left(\k-\k'\right) U_s+{U_c} \chi_c \left(\k-\k'\right) U_c\right]_{l_1l_2, l_3l_4} \quad &\text{for triplet}
\end{aligned}\right.
\]
We note that in Eq.~(\ref{eq:16}) $(l_1l_2,l_3l_4)$ are labels of atomic orbitals and $(ij,mn)$ are labels of band indices.

\subsection{Linearized Gap Equation}

We show the details of linearized gap equation calculations below. We focus on a two-dimensional Brillouin zone (which is relevant to the bilayer two-orbital model in this work).

The linearized gap equation for multi-orbital system is: 
\[\label{eq:18}
\Delta_i(\k) = - \sum_{j} \int_{FS_j}d\k'\frac{1}{|\textbf{v}_j(\k')|}\Gamma_{ij}(\k, \k')\Delta_j(\k')
\]
where $FS_j$ means that the integration is performed on the $j$-th Fermi surface sheet. $\textbf{v}_j(\k')$ is the Fermi velocity at the momentum $\k$ of $j$-th Fermi surface sheet. The effective pairing potential $\Gamma_{ij}(\k, \k') = V^{\rm{s}/\rm{t}}_{jj,ii}(\k, \k')$ (see Eqs.~(\ref{eq:15})-(\ref{eq:17})). $\Delta_i(\k)$ is the gap function of $i$-th Fermi surface sheet. With a discrete $\k$-mesh, Eq.~(\ref{eq:18}) is reduced to an eigenvalue problem:
\[\label{eq:19a}
\lambda \Delta_i(\k) = M_{ij}(\k,\k')
 \Delta_j(\k')
\]
where the matrix $M_{ij}(\k,\k')$ is:
\[\label{eq:20}
M_{ij}(\k,\k') = -\frac{1}{(2\pi)^2}\frac{l_j(\k')}{|\textbf{v}_j(\k')|}\Gamma_{ij}(\k,\k')
\]
and $l_{j}(\k')$ is the length element at $\k'$ grid of $j$-th Fermi surface sheet and $\textbf{v}_j(\k')$ is the Fermi velocity of the $j$-th energy band at $\k'$. The calculation of $l_{j}(\k')$ is shown in Supplementary Note 7. All momenta $\k$ are constrained close to the Fermi surface such that $|\epsilon_{\k} - \mu| \le \xi_c$ where $\mu$ is the chemical potential and $\xi_c$ is an energy cutoff. $\lambda$ is the eigenvalue, which is real. The largest eigenvalue corresponds to the leading superconducting instability. 

In this work, for the linearized gap equation calculations, we use a $\k$-mesh of $200\times 200\times1$ to sample the 2D square Brillouin zone and $\xi_c = $ 1 meV. The temperature $T$ is set at 116 K. In Wannier's model, we find 1576 $\k$ points on the Fermi surface. In Luo's model, we find 1436 $\k$ points on the Fermi surface.

\section*{Data availability}
The data that support the findings of this study are available from the corresponding author upon reasonable request. Source data are provided with this paper.

\section*{Code availability}
The electronic structure calculations were performed using the proprietary code VASP~\cite{PhysRevB.54.11169} and the open-source code Wannier90~\cite{MOSTOFI2008685}. Wannier90 is freely distributed on academic use under the Massachusetts Institute of Technology (MIT) License.

\begin{acknowledgments}
  We are grateful to Steffen B\"otzel, Frank Lechermann, Xianxin Wu, Fan Yang, Yi-feng Yang and Dao-Xin Yao for useful discussions. This project was financially supported by the National Natural Science Foundation of China under project number 12374064, the National Key R\&D Program of China under project number 2021YFE0107900, and Science and Technology Commission of Shanghai Municipality under grant number 23ZR1445400. NYU High-Performance-Computing (HPC) provides computational resources.
\end{acknowledgments}

\section*{Author contributions}
H.C. conceived and supervised the project. C.X., H.L., S.Z. and H.C. performed the calculations. H.C., C.X. and H.L. wrote the manuscript. All the authors participated in the discussion.

\section*{Competing interests}
The authors declare no competing financial and non-financial interests.

\bibliography{main}

\end{document}